\documentstyle[11pt,moriond,epsfig]{article}

\def\Journal#1#2#3#4{{#1} {\bf #2}, #3 (#4)}

\def\PLB{{\em Phys. Lett.}  B}
\def\PRL{\em Phys. Rev. Lett.}
\def\PR{\em Phys. Rev.}
\def\PRD{{\em Phys. Rev.} D}

\def\ra{\rightarrow}

\def\be{\begin{equation}}
\def\ee{\end{equation}}
\def\bea{\begin{eqnarray}}
\def\eea{\end{eqnarray}}

\def\ti         {\tilde}
\def\sel        {\ti{e}}
\def\smu        {\ti{\mu}}
\def\stau       {\ti{\tau}}   
\def\stop       {\ti{t}}
\def\sbot       {\ti{b}}  
\def\chino       {\ti{\chi}}
\def\gsim{\mathrel{\rlap {\raise.4ex\hbox{$>$}}{\lower.6ex\hbox{$\sim$}}}}
\def\lsim{\mathrel{\rlap {\raise.4ex\hbox{$<$}}{\lower.6ex\hbox{$\sim$}}}}    
\def\cost               {\cos\theta_{LR}}              

\newcommand{\chha}{$\tilde{\chi}_1^{\pm}$}
\newcommand{\chacha}{$\tilde{\chi}_1^+\tilde{\chi}_1^-$}
\newcommand{\chnchn}{$\tilde{\chi}_1^0\tilde{\chi}_2^0$}
\newcommand{\chna}{$\tilde{\chi}_1^0$}
\newcommand{\chnb}{$\tilde{\chi}_2^0$}

\newcommand{\snu}{$\tilde{\nu}$}

\def\sutwo{SU(2)_{L}}

\newcommand{\DM}{\Delta M}

\begin{document}

\vspace*{1.2cm}

\title{MSSM SUSY SEARCHES AT LEP2}

\author{IA IASHVILI}

\author{For the LEP Collaborations}

\address{DESY Zeuthen, Platanenallee 6, 15735, Germany \\ 
          and \\
Institute of Physics, Georgian Academy of Sciencies, Tbilisi}

\vspace*{2cm}

\maketitle\abstracts{
The status of SUSY searches at LEP2 up to centre-of-mass 
energy of $\sqrt{s}$=202\,GeV is presented. Search strategies for
sleptons, squarks, charginos and  neutralinos
are discussed in the framework of Minimal Supersymmetric Standard Model
with R-parity conservation.
With no indication for the production of these particles 
new limits are set on their masses.}

\newpage

\section{Introduction}

Increased energies and luminosities at LEP2 
have substantially extended its potential for
the discovery of new particles. In 1999 LEP 
operated at $\sqrt{s}=192-202\,$GeV and 
has delivered integrated luminosity
of 230\,pb$^{-1}$ per experiment.
This data, together with those  recorded during previous years
at $\sqrt{s} > M_Z$, have been used to
search for evidence for new physics beyond the Standard Model (SM).

Extensive searches have been 
performed for the particles predicted by Supersymmetric (SUSY) theories.
Among SUSY models Minimal Supersymmetric Standard Model
(MSSM)~\cite{mssm} 
with the unification of gaugino masses and scalar masses at GUT, 
often referred to
as Constrained MSSM (CMSSM)~\cite{cmssm}, is widely accepted as a main 
framework  at LEP. CMSSM introduces 6 new parameters:
$M_2$, the $\sutwo$ gaugino mass; 
$m_0$ and $A_0$,  a universal scalar mass and universal trilinear coupling  
at GUT scale; 
$\tan\beta$, the ratio of VEV's of the two Higgs fields;
$\mu$, the higgsino mass parameter and
$M_A$, the mass of CP-odd Higgs boson. 
Masses and couplings of SUSY particles (sparticles) as well as their production 
cross-sections are entirely determined once the first five parameters are
specified.


At LEP2 production of sleptons, $\ti{\ell}$, squarks, $\ti{q}$, 
charginos, $\ti{\chi}^{\pm}_i \ (i=1,2)$ and 
neutralinos, $\ti{\chi}^0_i \ (i=1,2,3,4)$,
all can take place with sizable cross-sections if kinematically allowed.
R-parity conservation implies that sparticles are produced
in pairs and ultimately decay to a stable Lightest Supersymmetric Particle
(LSP).
In CMSSM the best candidate for LSP is the
lightest neutralino $\ti{\chi}_1^0$. Since $\ti{\chi}_1^0$ is
weakly interacting it escapes detection and produces
energy imbalance in the event. 
At LEP2 energies sparticles would 
mainly decay directly to LSP. For sfermions the dominant
decay pattern is $\mathrm{\tilde{f} \to f \ti{\chi}_1^0}$,
except for stop, for which the relevant decay modes
are $\mathrm{\tilde{t} \to c \chino_1^0}$ and 
$\mathrm{\tilde{t} \to b \ell \ti{\nu}}$
since flavour conserving decay to top is 
kinematically unaccessible.
Charginos and neutralinos decay trough
$\mathrm{\ti{\chi}_1^+   \to f \bar{f}' \ti{\chi}_1^0}$ and
$\mathrm{\ti{\chi}_{i}^0 \to f \bar{f}  \ti{\chi}_1^0}$.
On the other hand, cascade decays such as
$\mathrm{\tilde{f} \to f \ti{\chi}_2^0 \to f f \bar{f} \ti{\chi}_1^0}$,
$\ti{\chi}_1^+ \to \tilde{\ell} \nu (\ell \tilde{\nu}) \to \ell \nu
\ti{\chi}_1^0$,
$\mathrm{\ti{\chi}_3^0 \to f \bar{f}  \ti{\chi}_2^0 \to 
f \bar{f} f \bar{f} \ti{\chi}_1^0}$,
$\mathrm{\ti{\chi}_i^0 \to \tilde{\ell} {\ell} \to \ell \ell
\ti{\chi}_1^0}$,
are also 
important in some regions of the model parameter space.
Depending on the identity of the fermion(s) produced in these
decays one expects lepton(s), or lepton(s) plus jets, or jets
in the final states,
all accompanied by missing energy as a fingerprint of the  escaping LSP's. 
Since at least two LSP's are present  no explicit reconstruction of
sparticle masses is possible. Thus the search strategy is to look
for the event excess over the SM expectations in the above 
listed final states.
The signal kinematics depends strongly on the mass
difference between produced sparticle and its invisible decay product,
$\DM=M_{spart}-M_{LSP}$, the quantity which essentially defines 
the amount of detectable energy.

There are number of SM processes taking place at LEP which may mimic
a signal. They can be classified as two-fermion production
f$\mathrm{\bar{f}(\gamma)}$, processes with four-fermions in the final
states, W$^+$W$^-$, We$\nu$, ZZ, Ze$^+$e$^-$ 
and two-photon interactions,
e$^+$e$^-\gamma\gamma \to \,$e$^+$e$^-$q$\mathrm{\bar{q}}$, e$^+$e$^-\ell^-\ell^+$.
When $\DM$ is low, in the range of 5\,GeV to 15--20\,GeV, 
the visible energy in SUSY signal events is low. 
Here the main and essentially
irreducible background is two-photon interactions
with final e$^+$/e$^-$
usually escaping undetected in the beam pipe and thus producing missing
energy. Furthermore, the cross-section of two-photon interactions 
($\sim$16~nb at $\sqrt{s}$=192--202\,GeV with $M_{\gamma\gamma}>3\,$GeV)
is significantly higher than that of any other processes at LEP2. Hence SUSY signal
with low $\DM$ is the most difficult to probe.
At high $\DM$ values, $\gsim$40--50\,GeV, 
the dominant backgrounds
are W$^+$W$^-$  and We$\nu$ productions.
In the intermediate $\DM$ range of 20\,GeV to 40\,GeV, the two-fermion,
four-fermion and two-photon processes all can contribute to the
background with comparable proportion. Apparently, 
dedicated selections are necessary 
to cover all possible ranges of $\DM$.
The key kinematical selection variables are visible energy and mass, missing mass, 
acollinearity, fraction of energy deposition in the forward
regions of the detector, event thrust, etc.
The selection techniques are conventional, or multivariate, which take into account correlation
between kinematical variables in the multidimensional space.                                                                       

Special situation arises when produced sparticle and its invisible decay product
are almost mass degenerate,
e.\,g. $\DM\lsim 3\,$GeV. The amount of 
detectable energy carried by decay products becomes too small and 
``standard'' search strategies are not applicable any more.
On the other hand, produced SUSY sparticles can become long-lived
due to decay phase-space suppression
and can be searched 
via displaced vertices (quasi-stable sparticle case) or anomalous
ionisation measurement (stable sparticle case).
Another technique is to tag events with Initial State Radiation (ISR)
photon(s).

\begin{figure}[t]
\hspace*{5mm}\begin{minipage}{0.45\textwidth}
\includegraphics[bb=25 25 550 600,
width=7.3cm,height=6.3cm,clip=true,draft=false]{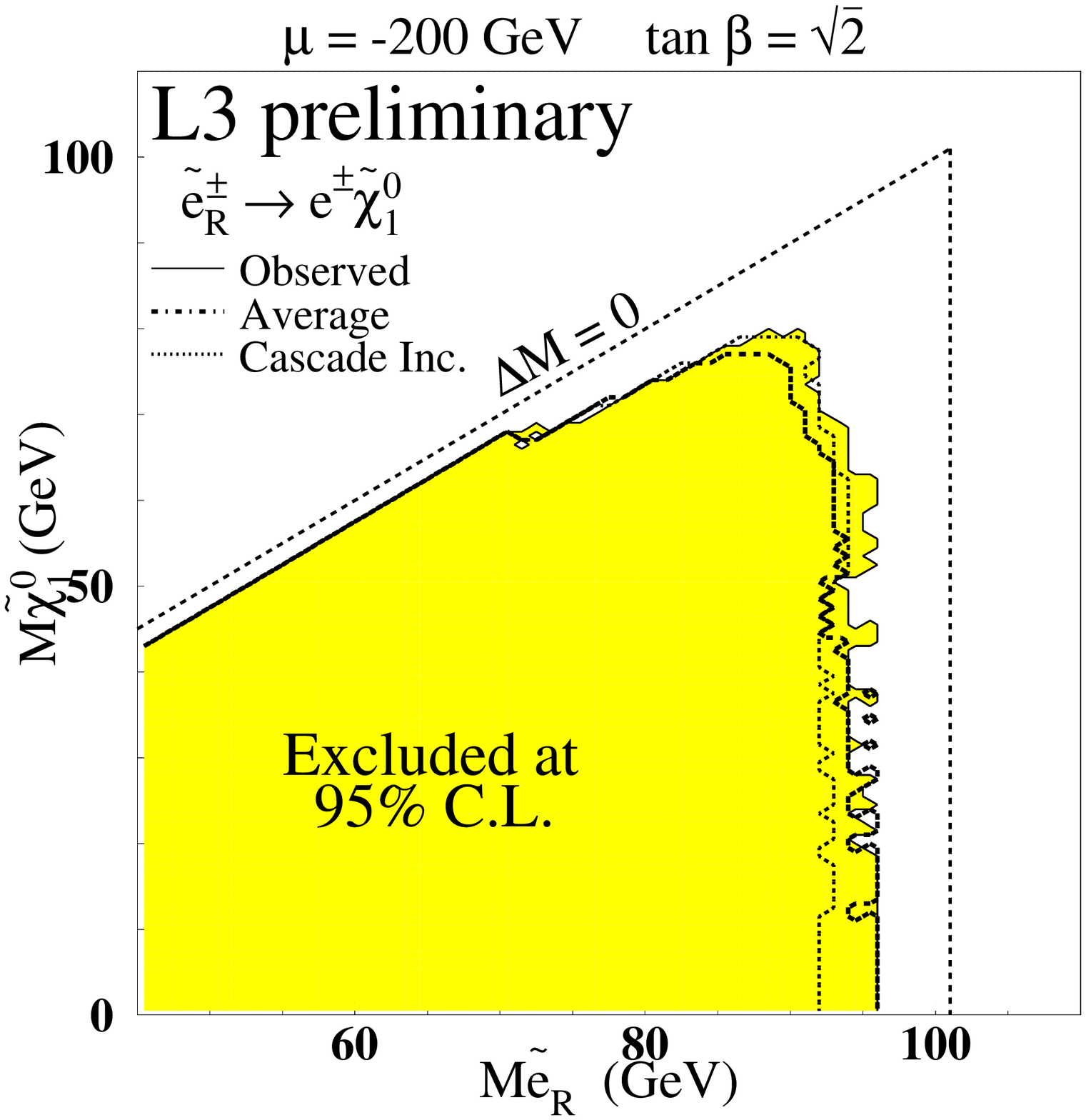}
\end{minipage}
\hspace*{-15mm}\begin{minipage}{0.45\textwidth}
\includegraphics[bb=35 172 526 655, 
width=7.3cm,height=5.5cm,clip=true,draft=false]{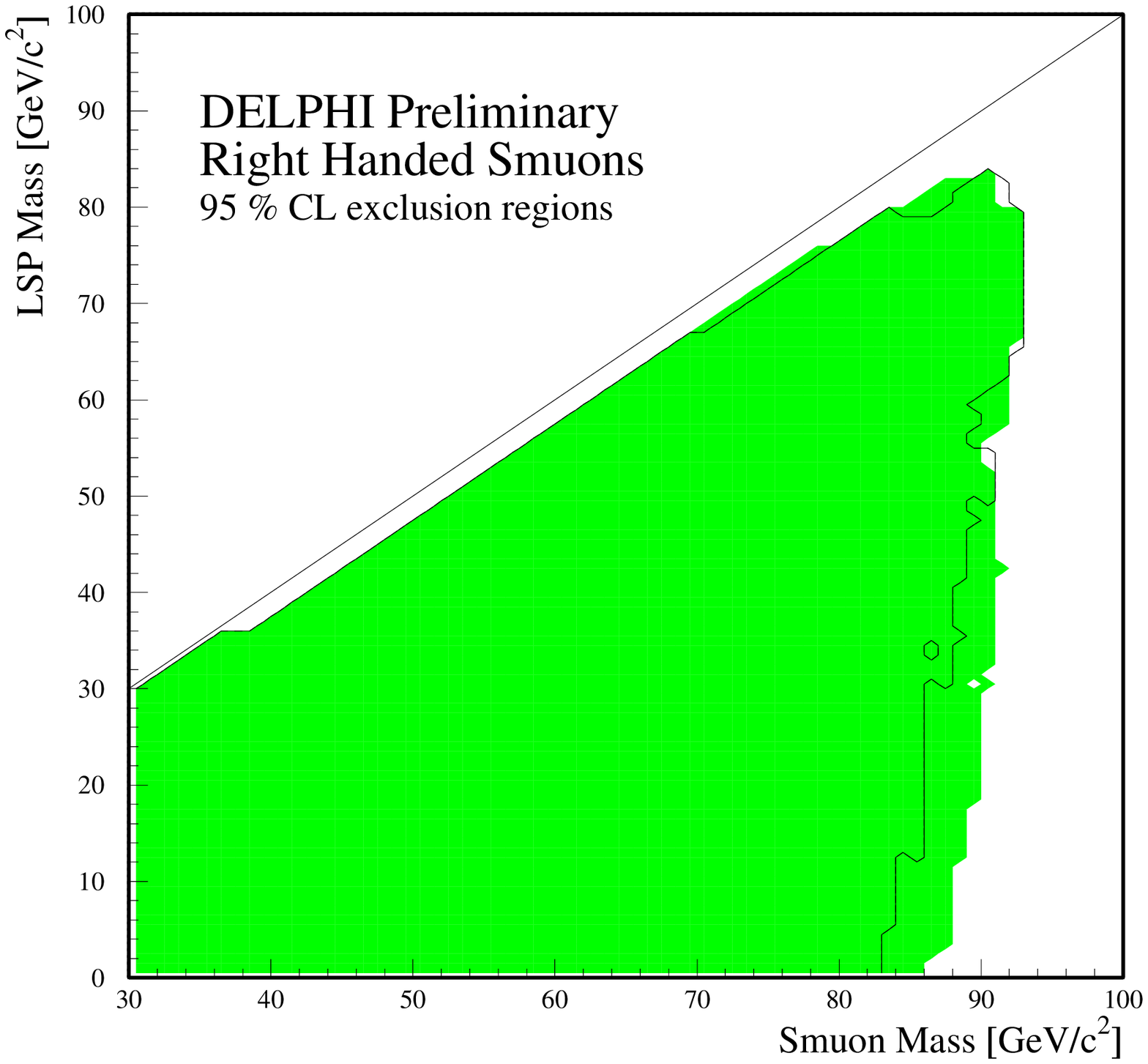}
\end{minipage}
\hspace*{2mm}\begin{minipage}{0.45\textwidth} 
\includegraphics[bb=25 25 350 350,width=9.0cm,height=6.2cm,clip=true,draft=false]{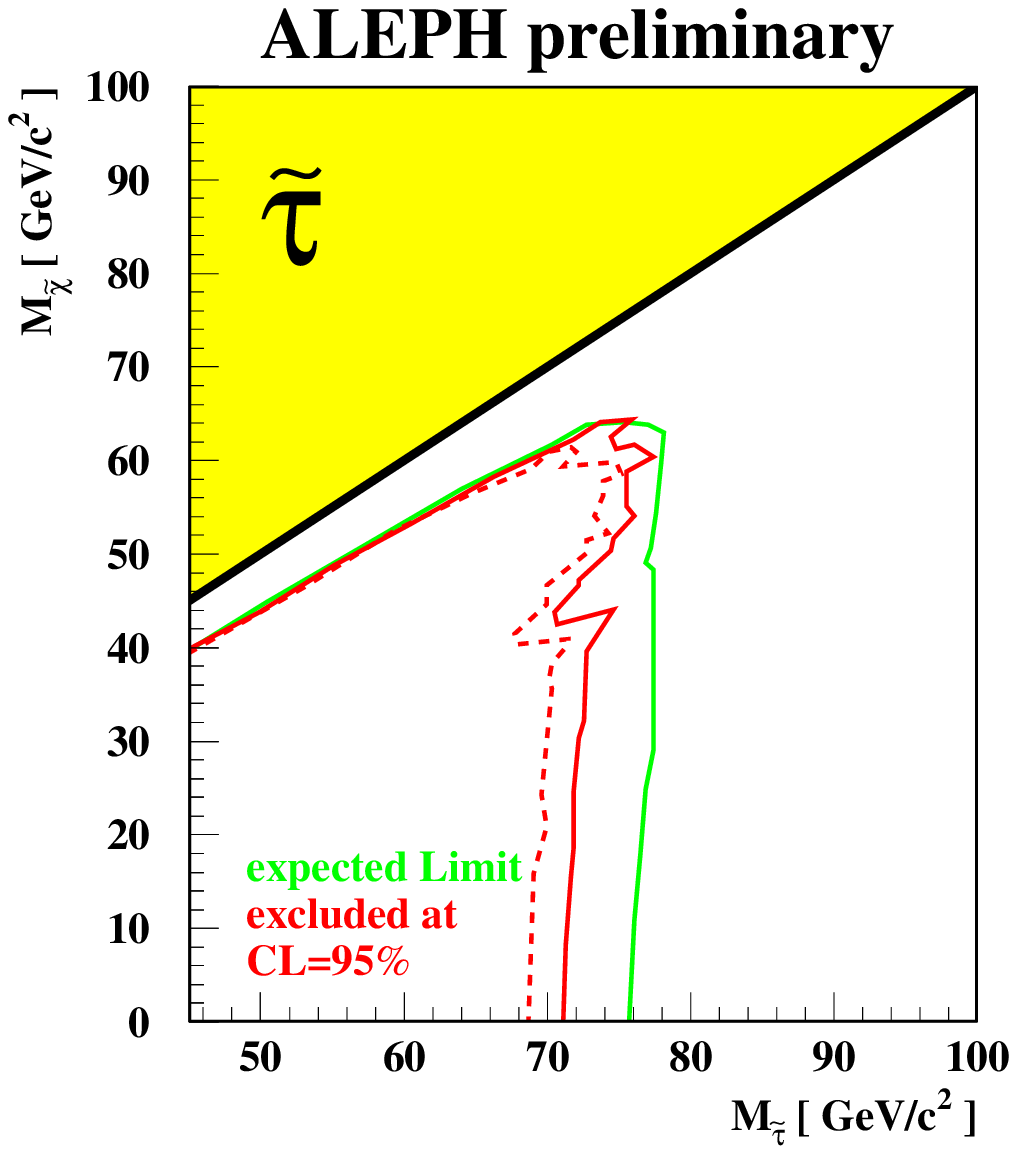} 
\end{minipage}
\hspace*{10mm}\begin{minipage}{0.45\textwidth}
\caption{95\% C.L. lower mass limits for: 
selectron (L3, top, left), smuon (DELPHI, top, right)
and stau (ALEPH, bottom). In case of $\mathrm{\ti{\mu}_R}$
limits are derived  for $\tan\beta=1.5$ and $\mu=-200\,$GeV
taking into account cascade decays; the
shaded region shows obtained exclusion limit and the
solid line shows expected limit.
In case of stau 100\% branching ratio is assumed for $\mathrm{\ti{\tau}
\to \tau}$ \chna \ decay;
dark and light solid lines show obtained and expected limits assuming no
mixing; dashed
line shows obtained limits for minimal cross-section 
when left--right mixing is  taken into account.} 
\end{minipage} 
\end{figure}

\section{Searches for SUSY Partners of Fermions}   

SUSY introduces scalar partners for each SM fermions, $\mathrm{\ti{f}_L}$
and $\mathrm{\ti{f}_R}$ which mix and form mass eigenstates,
$\mathrm{\ti{f}_1}$ and $\mathrm{\ti{f}_2}$.
The size of the mixing is proportional to the corresponding SM fermion
mass and can be neglected for the first two generations.
However, the stop left -- right mixing
is expected to be large due to heavy top quark.
Resulting eigenstates have sizeable mass splitting with the lighter
stop, $\mathrm{\ti{t}_1}$, being the lightest of all squarks. In case of
sbottom and stau
large mixing can occur for $\tan\beta \gsim 10$.

\begin{table}[t]
\label{tab:sleptons}
\caption{95\% C.L. lower mass limits for sleptons obtained by LEP experiments with high energy 
data up to $\sqrt{s}=192-202\,$GeV.}
\vspace*{2mm}
\hspace*{9mm}\begin{tabular}{|l|l|l|l|}\hline
Experiment 
& \ \ \ \ \ \ selectron 
& \ \ \ \ \ \ \ \ smuon 
& \ \ \ \ \ \ \ \ \ \ \ stau \\ \hline
ALEPH  
& $M_{\sel_R}>$92~GeV & $M_{\smu_R}>$85~GeV & $M_{\stau_R}>$70~GeV  \\
& for $\DM>$10~GeV    & for $\DM>$10~GeV    & for $\DM>$10~GeV  \\ 
& $\mu=-$200~GeV, $\tan\beta$=2
& BR($\smu_R\to\mu$\chna)=100\%
& BR($\stau_R\to\tau$\chna)=100\%
\\
\hline
DELPHI 
& $M_{\sel_R}>$91~GeV & $M_{\smu_R}>86$~GeV & $M_{\stau_R}>$75.5~GeV  \\
& for $\DM>$15~GeV
& for $\DM>$10~GeV
& for $\DM>$15~GeV
\\
& $\mu=-$200~GeV, $\tan\beta$=1.5
& $\mu=-$200~GeV, $\tan\beta$=1.5
& BR($\stau_R\to\tau$\chna)=100\%
\\
\hline 
L3  
& $M_{\sel_R}>$91~GeV & $M_{\smu_R}>$78~GeV & $M_{\stau_R}>$68~GeV  \\
& for $\DM>$15~GeV  
& for $\DM>$10~GeV
& for $\DM>$15~GeV  \\
& $\mu$=-200~GeV, $\tan\beta=\sqrt{2}$
& BR($\smu_R\to\mu$\chna)=100\%
& BR($\stau_R\to\tau$\chna)=100\% \\
\hline
\end{tabular}
\end{table}

\begin{figure}[t]
\hspace*{-4mm}\begin{minipage}{0.45\textwidth}
\includegraphics[bb=10 10 550  500,
width=7.02cm,height=6.00cm,clip=true,draft=false]{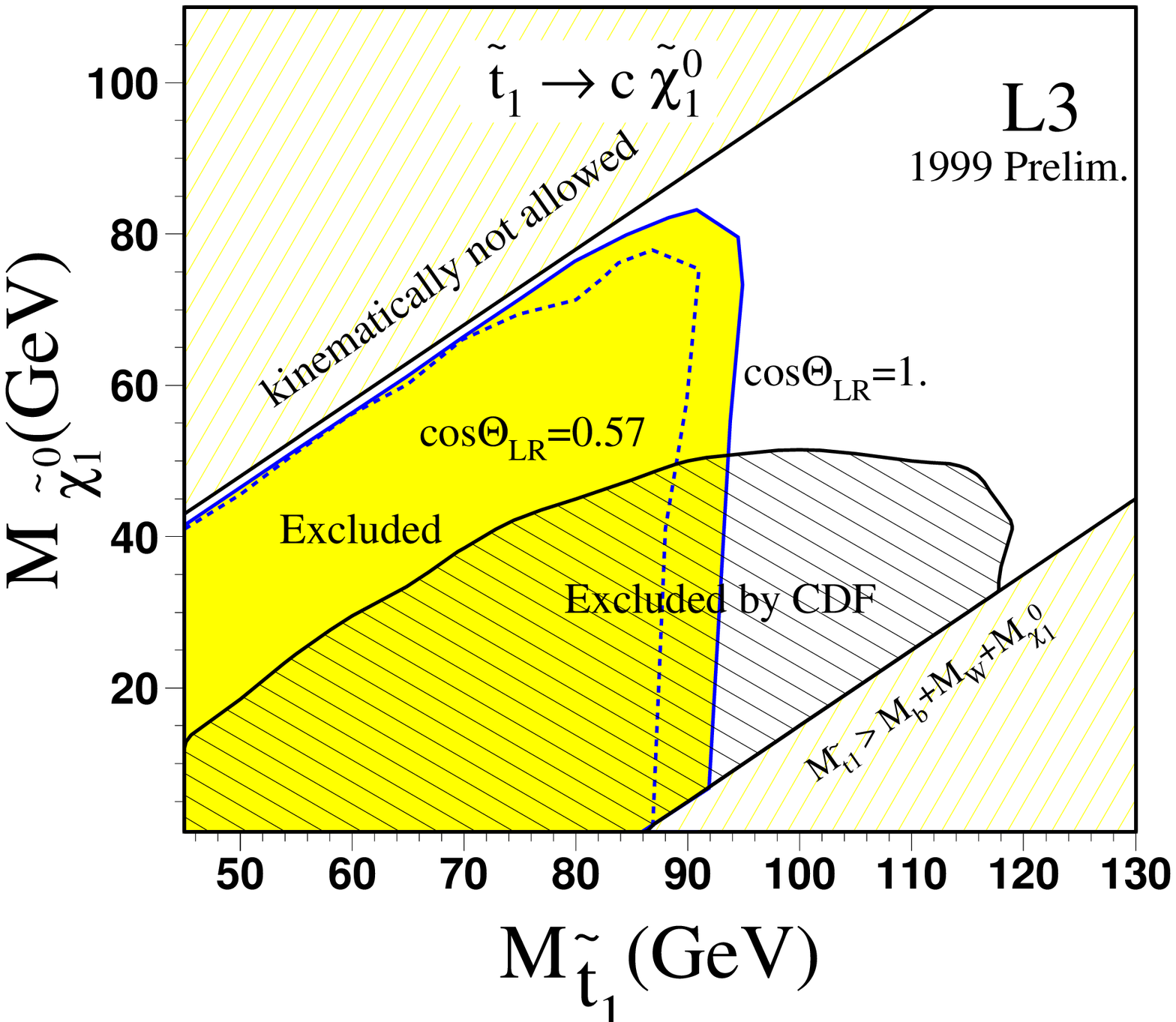}             
\end{minipage}
\begin{minipage}{0.45\textwidth}
\hspace*{5mm}\includegraphics[bb=10 10 410 443,
width=7.2cm,height=6.52cm,clip=true,draft=false]{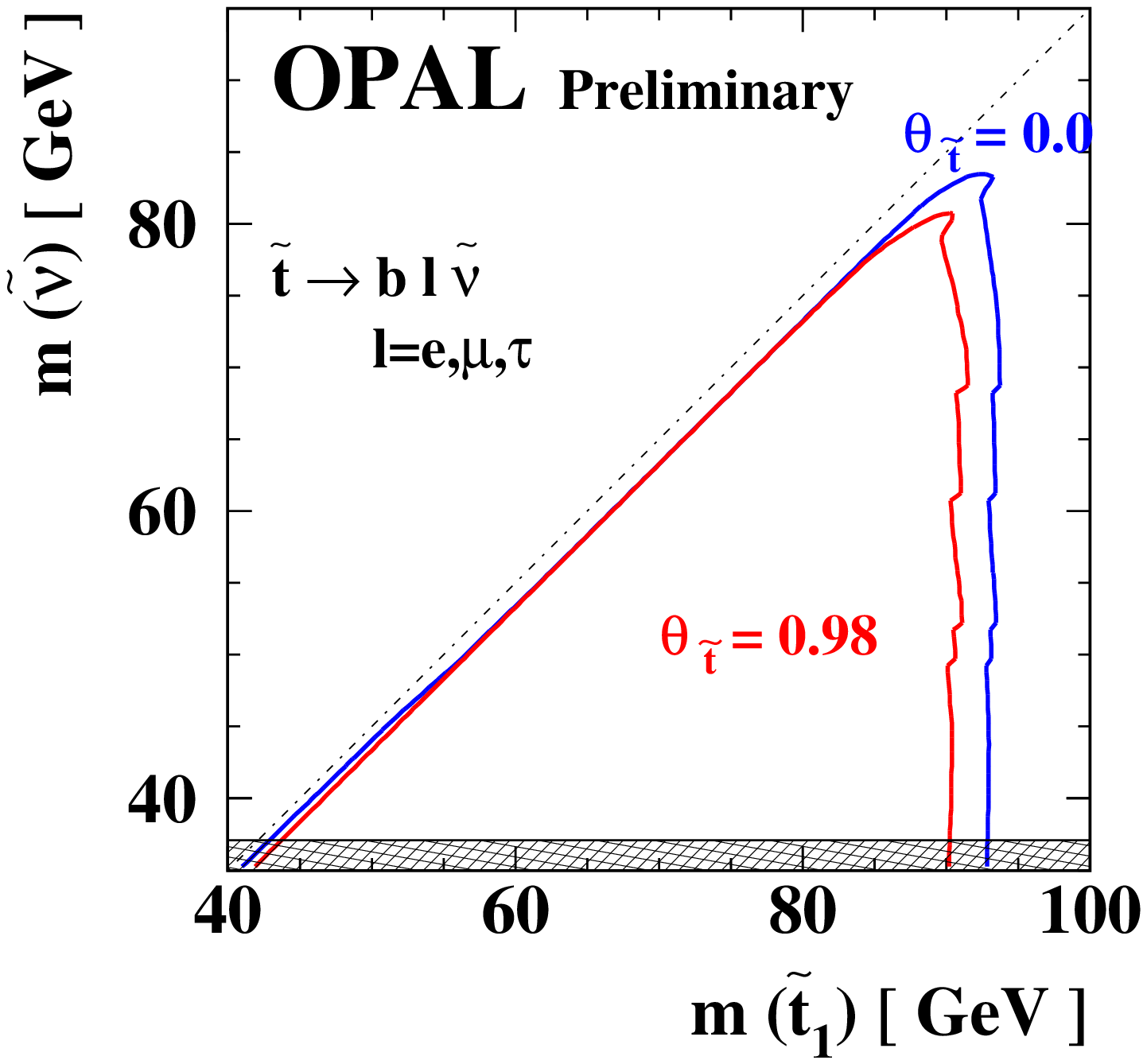}
\end{minipage}

\begin{minipage}{0.45\textwidth}
\includegraphics[bb=10 10 510 510,
width=6.57cm,height=5.25cm,clip=true,draft=false]{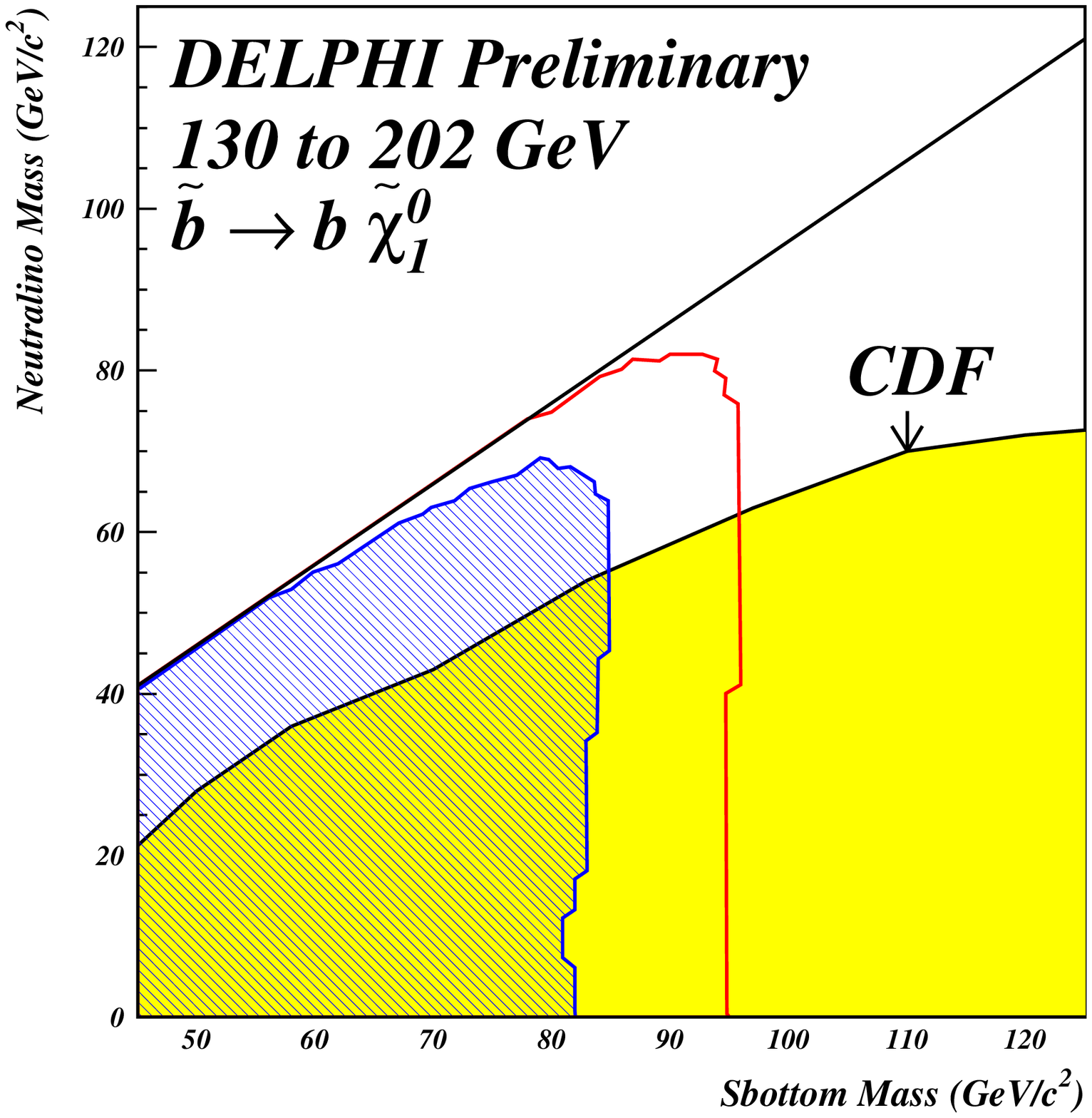}         
\end{minipage}
\begin{minipage}{0.45\textwidth}
\hspace*{5mm}\includegraphics[bb=10 10 510 510,
width=7.59cm,height=5.88cm,clip=true,draft=false]{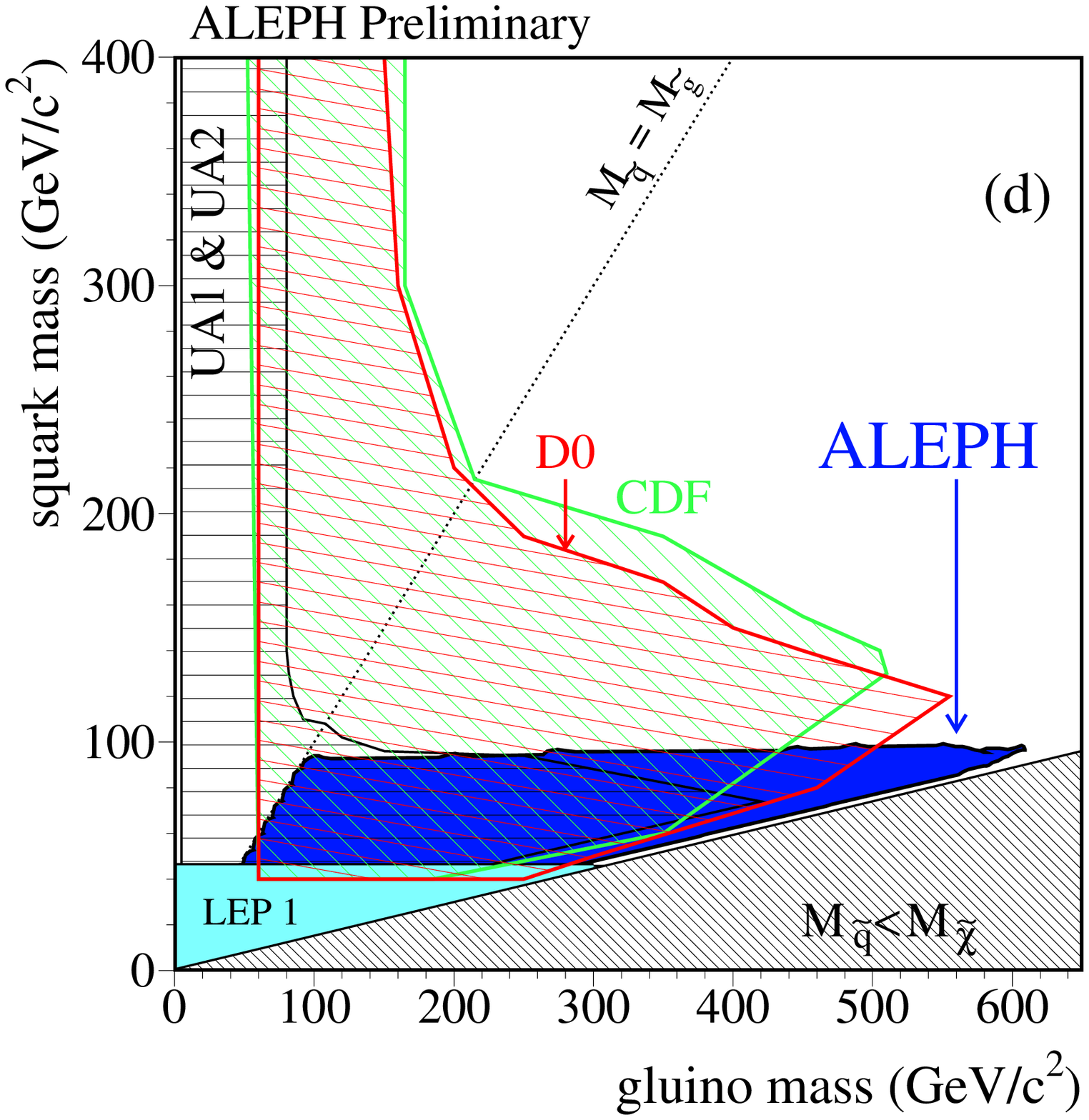}
\end{minipage}
\caption{95\% C.L. lower mass limits for:  stop decaying via 
$\mathrm{\ti{t}_1 \ra}$~c~$\ti{\chi}_1^0$ (L3, top, left);
stop decaying via  
$\mathrm{\ti{t}_1 \ra}$~b~$\ell \ti{\chi}_1^0$ (OPAL, top, right);
sbottom 
$\mathrm{\ti{b}_1 \ra}$~b~$\ti{\chi}_1^0$ (DELPHI, bottom, left)
and mass-degenerate squarks  
$\mathrm{\ti{q} \ra}$~q~$\ti{\chi}_1^0$,
$\mathrm{\ti{q}=\ti{u}, \ti{d}, \ti{c}, \ti{s}, \ti{b}}$
(ALEPH, bottom, right). The latter is shown
in ($M_{\ti{g}}$, $M_{\ti{g}}$) plane along with the regions excluded
by UA1~\protect\cite{ua1}, UA2~\protect\cite{ua2} and
Tevatron~\protect\cite{cdf_etmis+jets,d0_etmis+34jets} experiments.
The $\mathrm{\ti{t}_1}$ and  $\mathrm{\ti{b}_1}$
mass limits are given for maximal and
minimal cross-sections. The stop and sbottom mass limits obtained by
CDF~\protect\cite{cdf_squarks} are also shown.}
\end{figure}

Pair production of sfermions
in e$^+$e$^-$ collisions takes place through s-channel $\gamma$/Z
exchange. For selectrons the production cross-section is enhanced
by t-channel exchange of neutralino. Since right handed sleptons are
expected to be lighter than the left-handed states, it is
assumed that only $\ti{\ell}_R^{+}\ti{\ell}_R^{-}$ production 
is kinematically possible. The decay pattern $\ti{\ell} \to
\ell$ \chna \ gives rise
to two acoplanar leptons and missing energy in the final states. 
In this category of events no excess has been found
over the SM background 
expectations~\cite{aleph,delphi_sleptons,l3_sfermions,opal}. 
As an example,  L3 has observed 76 events
with two acoplanar electrons and missing energy 
in $\sqrt{s}=$192-202\,GeV data sample. This is consistent with 68.9
events expected from
the SM processes. Corresponding exclusion contour for the selectron 
mass is shown in Fig.~1 assuming
$\mu=-$200\,GeV and $\tan\beta=\sqrt{2}$.  The exclusion holds also
for higher values of $|\mu|$ and $\tan\beta$. For smaller values of
$|\mu|$ the  t-channel contribution to the selectron
pair production is suppressed yielding a few GeV decrease in
the limit. The values of $\mu$ and $\tan\beta$ are also
relevant for calculation of  branching ratio for 
cascade decays $\mathrm{\ti{e}} \to$ 
e \chnb $\to$ e f $\mathrm{\bar{f}}$ \chna.
The plot gives exclusion contour assuming 
BR($\mathrm{\ti{e}} \to $e\,\chna)=100\%; it also shows the
change in the exclusion contour due to opening up the cascade decays
for which vanishing efficiencies are (pessimistically) assumed.

Smuon pair production gives rise to two acoplanar muons and missing
energy. In this category of events 13 candidates have been observed by
DELPHI. This is consistent with 19.2 events expected from the SM
processes. The corresponding
exclusion plot is shown in Fig.~1 for $\tan\beta=1.5$ and
$\mu=-200$\,GeV
taking into account efficiency loss due to cascade decays.

In the search for stau pair production which leads to two taus and missing energy,
 46 events have been observed
by ALEPH, whereas 34.2 is expected from the SM processes. 
The excluded regions with the assumption of 
BR($\stau \to\tau$~\chna)=100\% is shown in Fig.~1. Exclusion contours are 
derived assuming no mixing and sizable mixing between the left and right
eigenstates.
In case of mixing the production cross-section can be parametrised as a
function of the sfermion mass and the
left-right mixing angle, $\cost$. The cross-section is maximal for
$\cost$=1
and is minimal for $\cost\simeq0.61$ when stau
decouples from the Z.
The contour shown 
for the mixing case 
corresponds to the minimal $\sigma(\ti{\tau}^+\ti{\tau}^-)$.

Table~1 summarises exclusion limits on slepton masses
obtained by ALEPH, DELPHI and L3 Collaborations. Individual limits 
on selectron mass are set at 91--92\,GeV 
for the mass differences of $\DM>$10--15\,GeV. Smuon mass exclusion ranges
from 78\,GeV to
85\,GeV for $\DM>$\,10\,GeV, whereas stau limits are relatively modest, 
in the range of 68--75.5\,GeV for $\DM>$10--15\,GeV.

Among squarks special emphasises are put on the stop and sbottom 
searches~\cite{aleph,l3_sfermions,opal,delphi_squarks} as
they are
expected to be lighter than others. The stop two-body decay 
$\mathrm{\tilde{t}_1 \to c \ti{\chi}_1^0}$ 
gives rise to final states with two jets and missing energy, whereas
three-body
decay mode  $\mathrm{\tilde{t}_1 \to b \ell \ti{\nu}}$ leads to two 
b-jets plus two leptons 
plus missing energy signature.
For sbottom, the 
$\mathrm{\tilde{b}_1 \to b \chino_1^0}$ decay results in two b-jets and 
missing energy final states. 
No indication for the $\mathrm{\tilde{t}_1}$
or $\mathrm{\tilde{b}_1}$
production has been observed 
in the above final states
and new limits have been derived on their masses, shown in Fig.~2. 
Exclusions are given
for maximal and minimal cross-section assumptions. Table~2 summarises 
mass limits for the pessimistic case of minimal cross-sections. 
The stop mass limits range
from 82\,GeV to 91.5\,GeV for the mass differences of $\DM>$10--15\,GeV, 
while the sbottom 
is excluded up to the masses of 76--81.5\,GeV for the same $\DM$ range.

Negative results of the search for the stop two-body decay
can be reinterpreted as a lower limit on the mass of 
$\mathrm{\ti{q}=\ti{u}, \ti{d}, \ti{c}, \ti{s}}$ ($\mathrm{\ti{b}})$
squarks, as these lead to a similar experimental signature. The
obtained results are presented in ($M_{\ti{q}}, M_{\ti{g}}$) plane,
Fig.~2, in order to directly compare with the Tevatron results.
Interpretation of the results in ($M_{\ti{q}}, M_{\ti{g}}$)  plane 
is possible thanks to gaugino 
masses unification assumption at GUT which relates
$M_{\ti{g}}$ and $M_{\ti{\chi}_1^0}$ at EW scale. The presented limit is
valid assuming mass degeneration between 
$\mathrm{\ti{u}, \ti{d}, \ti{c}, \ti{s}}$ and $\mathrm{\ti{b}}$, as well as
between their left and right eigenstates.

Dedicated searches have been performed for a stop almost mass
degenerate with LSP by ALEPH~\cite{aleph_stop}.
When  $\DM$
is less than the c-quark mass the stop decays through 
$\mathrm{\ti{t}_1 \to u}$\chna. The decay width for 
this mode  is much smaller
than for $\mathrm{\ti{t}_1 \to c}$~\chna \ which itself has 
width larger than typical hadronization scale. So for $\DM\lsim3$\,GeV
the decay length ranges from few microns to hundreds of meters depending
on SUSY model parameters. When stop decays outside detector it can be searched as
a heavy stable particle trough anomalous ionisation measurements. When decay occurs
inside detector the signal events can contain tracks with large
impact parameter.
Figure~3 shows complementarity of ``stable'', ``quasi-stable'' and ``standard'' analysis
in different $\DM$ regions. Assuming CMSSM relations  between SUSY 
parameters and $\Gamma(\ti{t}_1)$
the lower limit of $M_{\ti{t}_1}>63$\,GeV is set independent of
$\DM$.

\begin{table}[h] 
\caption{95\% C.L. lower limits on stop and sbottom masses for minimal cross-section
assumptions obtained by LEP experiments with
high energy data up to $\sqrt{s}=192-202$\,GeV.} 
\vspace*{2mm}
\hspace*{25mm}\begin{tabular}{|l|l|l|l|}\hline Experiment 
& \ \ \ \ \ \ \ stop & \ \ \ \ \ \ \ stop & \ \ \ \ \ \ \ sbottom
\\ \hline
ALEPH  
& $M_{\stop_1}>$87~GeV & $M_{\stop_1}>$88~GeV & $M_{\sbot_1}>$76~GeV
\\
& for 6$<\DM<$40~GeV     & for $\DM>$10~GeV   & for $\DM>$10~GeV  \\ 
\hline
DELPHI 
& $M_{\stop_1}>$91~GeV & $M_{\stop_1}>$82~GeV & $M_{\sbot_1}>$81~GeV  \\
& for $\DM>$15~GeV     & for $\DM>$15~GeV    & for $\DM>$15~GeV  \\
\hline
L3
& $M_{\stop_1}>$87~GeV 
& $M_{\stop_1}>$91.5~GeV 
& $M_{\sbot_1}>$80~GeV
\\
& for $\DM>$15~GeV     & for $\DM>$15~GeV    & for $\DM>$15~GeV  \\
\hline
OPAL
& $M_{\stop_1}>$91.5~GeV & $M_{\stop_1}>$89.5~GeV & $M_{\sbot_1}>$81.5~GeV
\\
& for $\DM>$10~GeV     & for $\DM>$10~GeV    & for $\DM>$10~GeV  \\
\hline
\end{tabular}
\end{table}

\begin{figure}[h]
\hspace*{40mm}
\includegraphics[bb=10 10 550 550,
width=9.2cm,height=7.2cm,clip=true,draft=false]{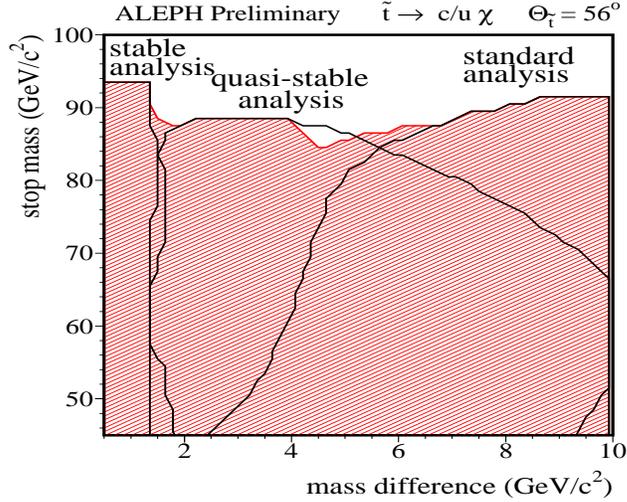}
\caption{95\% C.L. lower limits on the stop mass for $\mu=-100$~GeV,
$\tan\beta$=1.5 and $\theta_{LR}=56^{\circ}$ as a function of $\DM$
in the region of small mass differences. 
Regions of different applicable analysis are indicated.}
\end{figure}

\section{Searches for Charginos and Neutralinos}
In e$^+$e$^-$ collisions chargino pairs
$\tilde{\chi}^{+}_i\tilde{\chi}^{-}_j$ can be produced
through s-channel Z/$\gamma$ or t-channel
sneutrino exchange. Neutralino
pairs
$\tilde{\chi}^0_i\tilde{\chi}^0_j$ are produced
via s-channel Z or t-channel selectron 
exchange.
The cross-sections 
depend, apart from masses, on the Higgsino-gaugino contents
of the produced
sparticles. When $M_2\gg|\mu|$ the  \chha \ and 
$\tilde{\chi}^0_{1,2}$
are mostly Higgsinos and the t-channel
contributions
are negligible. For $M_2\ll|\mu|$ the \chha \ and
$\tilde{\chi}^0_{1,2}$ are gaugino-like and, if sfermions
are light, the t-channel contributions become important. 
In case of
\chacha \ production  the t-channel interferes
destructively with the s-channel leading to a significant
decrease of the 
$\sigma$(\chacha). On the contrary, the t-channel contribution enhances
the $\sigma$(\chnchn). 
At $\sqrt{s}=192-202$\,GeV in most of the CMSSM parameter space the
$\sigma$(\chacha) \ amounts to
a few pb for the chargino masses up to almost the kinematic limit.
For some sets of CMSSM parameters $\sigma$(\chnchn) can
also reach a few pb. 

\begin{figure}[t]
\begin{minipage}[b]{0.45\textwidth}
\hspace*{0mm}\includegraphics[bb=10 280 560 525,width=8.cm,height=6.cm,
clip=true,draft=false]{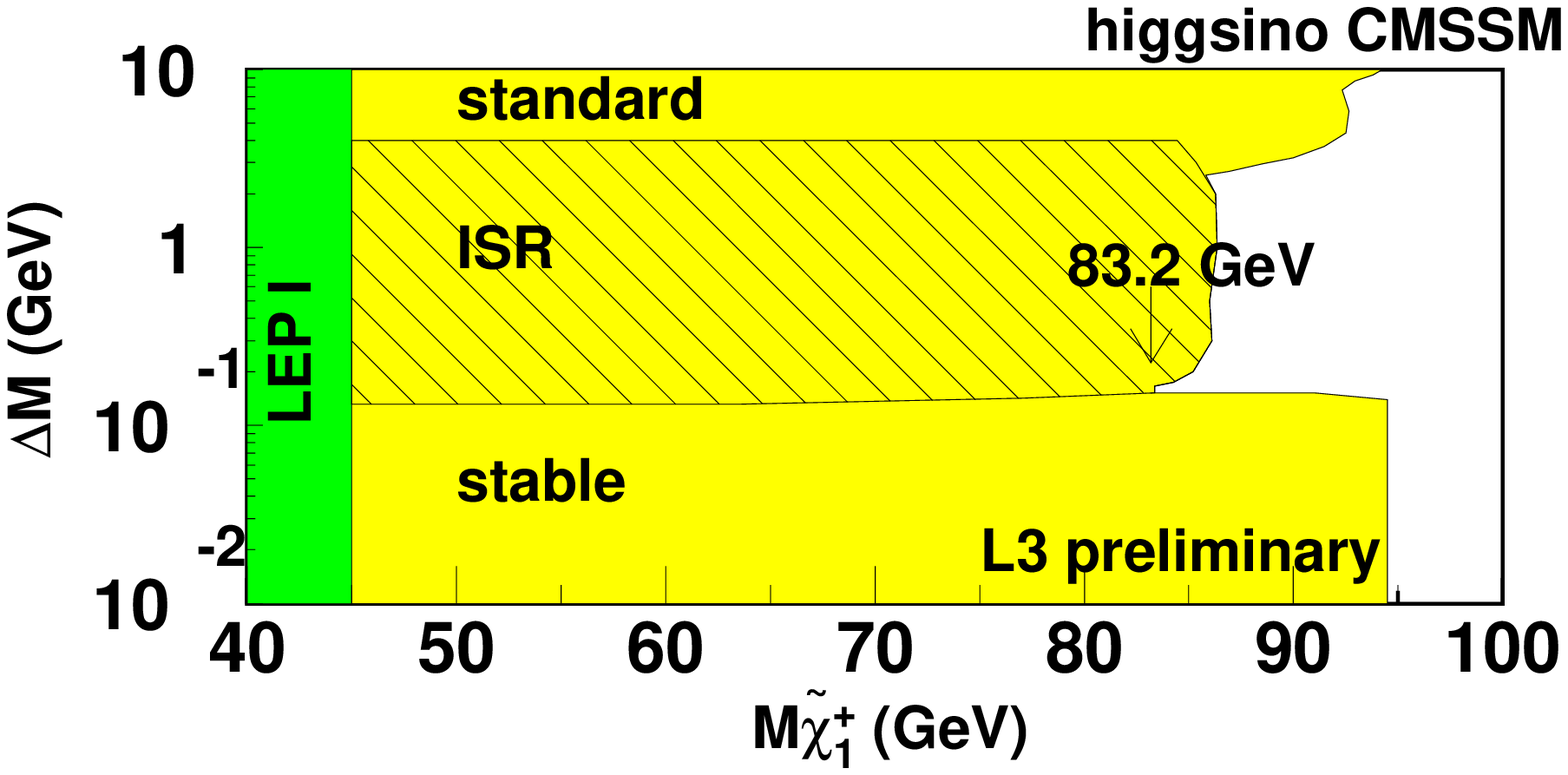} 
\caption{95\% C.L. lower limit on chargino mass as a function of $\DM$. Regions excluded 
from ``standard'', ISR and stable particle searches are indicated. 
}
\end{minipage}\hspace*{7mm}
\begin{minipage}[b]{0.45\textwidth}
\hspace*{0mm}\includegraphics[bb=30 10 550 550,width=9.2cm,height=7.0cm,
clip=true,draft=false]{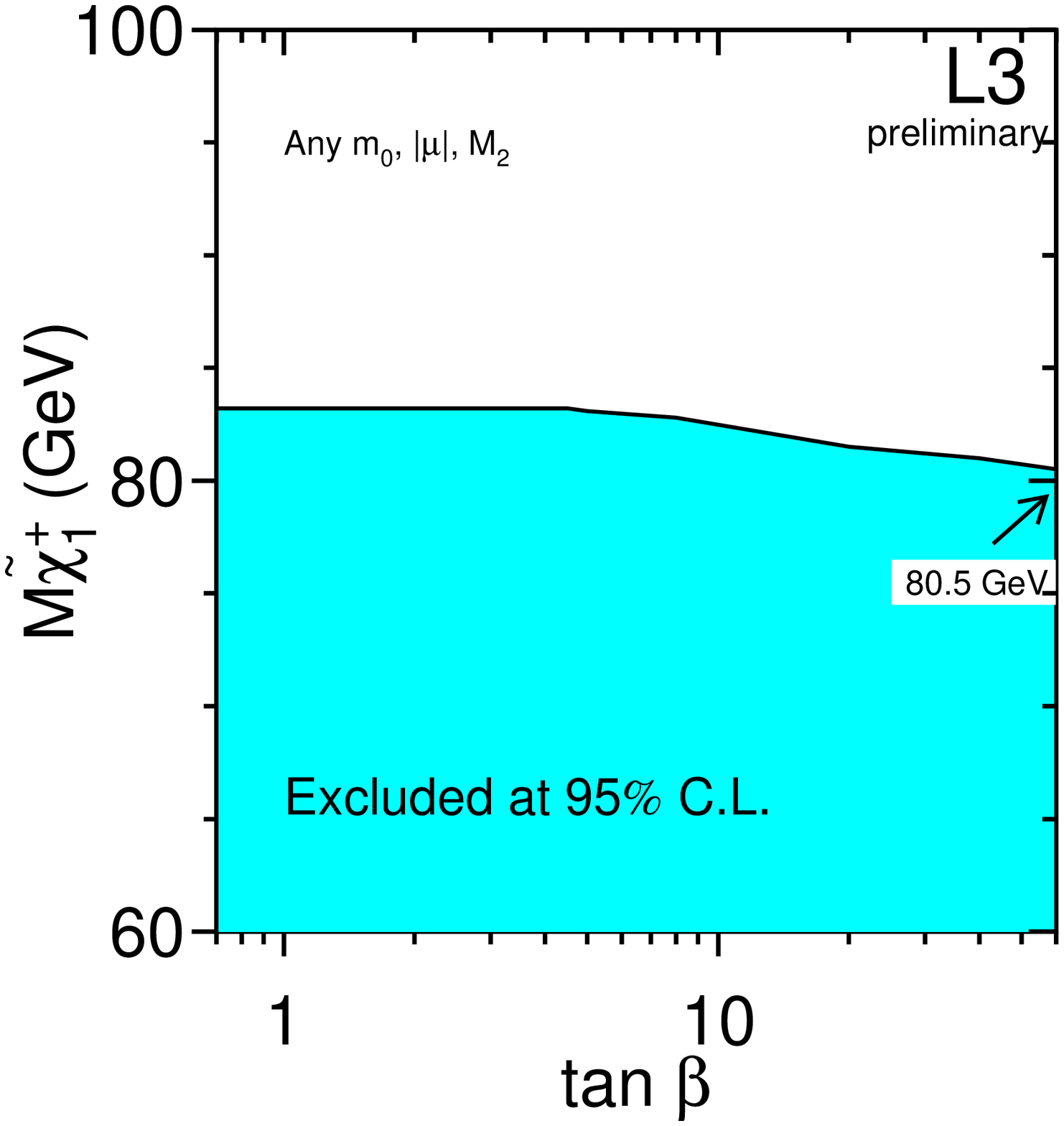}
\caption{95\% C.L. lower limit on chargino mass as a function of $\tan\beta$. Other CMSSM 
parameters are scanned in the range relevant for EW scale SUSY.}
\end{minipage}
\end{figure}

The expected event topologies from chargino and neutralino productions are 
jets plus missing energy, leptons plus
missing  energy or jets plus lepton(s) plus missing  energy. 
Searches through these topologies have been carried out by LEP 
experiments~\cite{aleph,opal,delphi_inos,l3_inos} at
high energies up to $\sqrt{s}=192-202$\,GeV.  
No evidence has been
found for the chargino or neutralino
production and limits are derived on their masses. 
For example, assuming $\tan\beta=1.5\,$(35),
$m_0>$500\,GeV and $\DM\ge10\,$GeV, the
mass limits of $M_{\ti{\chi}_1^{\pm}}>$100.0\,GeV
(100.1\,GeV), $M_{\ti{\chi}_2^0}>$75.4\,GeV (100.1\,GeV), and 
$M_{\ti{\chi}_3^0}>$110.9\,GeV (129.4\,GeV) have been set by OPAL
employing searches for $\ti{\chi}_1^+\ti{\chi}_1^-$, $\ti{\chi}_1^0\ti{\chi}_2^0$, 
$\ti{\chi}_2^0\ti{\chi}_2^0$ and
$\ti{\chi}_1^0\ti{\chi}_3^0$ productions. 

Dedicated search strategy has been devised for the chargino
mass degenerate with its invisible decay product,
$\DM\lsim4\,$GeV~\cite{l3_inos,delphi_chargino_degen}. 
In CMSSM this scenario may
occur is \chha \ if higgsino-like. Here, chargino
events can be tagged, e.g. with ISR  photon(s),
so the corresponding final state contains photon(s), missing energy
and soft tracks.
While in signal events 
the missing energy is due to weakly
interacting particles, in the dominant two-photon background it is due to 
final state e$^+$/e$^-$ escaping in the beam pipe. In the latter case,
if a
photon is present satisfying
$E_T^\gamma\ge\sqrt{s}\frac{\sin\theta_d}{1+\sin\theta_d}$ requirement,
the e$^+$ and/or e$^-$ must be deflected into detector and the event
can be identified as
a background. Here $\theta_d$ is the minimum detection angle 
for the deflected e$^+$/e$^-$.
For the L3 detector $\theta_d=1.5^{\circ}$  leads to $E_T^\gamma \ge
5.5$\,GeV ---
an energy large enough for efficient triggering.
The obtained mass limit 
is shown in Fig.~4 together with the exclusion contours obtained
by the ``standard'' searches and by stable charged particle searches.
These different
search strategies allow to close up the low $\DM$ ``window''.
The absolute mass limit for the \chha \ derived by L3 is shown
in Fig.~5 as a function
of $\tan\beta$. The limit is valid for any chargino field content and for any value of $m_0$.
The model parameters are
scanned in the entire range relevant for the EW scale SUSY:
0.7~$\le \tan\beta \le$~70, $M_2 \le 2000\,$GeV,
$|\mu|\le2000\,$GeV,  $m_0 \le 2000\,$GeV.
At low $\tan\beta$ values the limit on the chargino mass
is obtained when 
\chha \ and \chna \ are mass degenerate (higgsino region).
At large $\tan\beta$ values, on the other hand, 
the limit is obtained when the
\chha \ and \snu \ are mass degenerate (gaugino region).
The absolute limit is set at $M_{\ti{\chi}^{\pm}_1}>80.5\,$GeV.

The mass limit on \chna--LSP can be derived only indirectly
through searches of other SUSY particles. 
The $M_{\ti{\chi}_1^0}$ limits from LEP experiments are summarised in Tab.~3.
For large values of $m_0$, which implies heavy sfermions, the limit is derived using
results of chargino and neutralino searches. The LSP mass limits obtained by DELPHI and OPAL
assuming high values of $m_0$ are 35.2\,GeV and 35.7\,GeV, respectively.
For low values of $m_0$ sfermion searches become
important, whereas contributions from chargino pair production diminishes if
\chha \ is gaugino-like. 
General (independent of $m_0$) \chna \ mass
limits~\cite{l3_inos,aleph_lsp} are derived by 
L3 and ALEPH and are shown
in Fig.~6 as a function of $\tan\beta$.  
In case of L3 the lowest mass limit of 37.5\,GeV is observed at
$\tan\beta=1$ and high $m_0=500\,$GeV and
comes from searches of chargino and neutralino pairs.
For large $\tan\beta$ values
the minimum is found in the gaugino region and for low values of $m_0$. 
Here the limit is derived from the slepton searches.

The LSP mass limit derived by ALEPH exploits, along with sparticle searches,
lower limit on the light Higgs boson mass 
set at $M_h>$107.7\,GeV for low values of $\tan\beta$~\cite{aleph_lsp}. As
indicated in 
Fig.~6 the Higgs search constraints are important for the
regions of $\tan\beta\lsim2$ and
$3\lsim\tan\beta\lsim3.5$. 
Slepton searches allow to set limit 
at high values
of  $\tan\beta$ and chargino searches contribute in the region of 
 $2\lsim\tan\beta\lsim 3$. An absolute limit is set at
$M_{\ti{\chi}_1^0}>38$\,GeV.
The results obtained using Higgs
searches are however quite sensitive to the top
quark mass which is assumed to be $M_t$=175\,GeV~\cite{aleph_lsp}.

\begin{table}[h]
\caption{95\% C.L. lower limits on LSP masses obtained by LEP experiments with high energy
data up to $\sqrt{s}=192-202$~GeV.}                                            \vspace*{2mm}
\hspace*{30mm}\begin{tabular}{|l|ll|}\hline
ALEPH  & $M_{\ti{\chi}_1^0}>$38~GeV   & for $M_0\le$1~TeV, $M_t$=175~GeV
\\ \hline
DELPHI & $M_{\ti{\chi}_1^0}>$35.2~GeV & for $M_{\ti{\nu}}>$300~GeV  
\\ \hline
L3     & $M_{\ti{\chi}_1^0}>$37.5~GeV & 
\\ \hline
OPAL   & $M_{\ti{\chi}_1^0}>$35.7~GeV & for $m_0\ge500$~GeV 
\\ \hline
\end{tabular} 
\end{table}

\begin{figure}[t] 
\begin{minipage}{0.45\textwidth} \includegraphics[bb=10 10 550 587,
width=8.1cm,height=7.2cm,clip=true,draft=false]{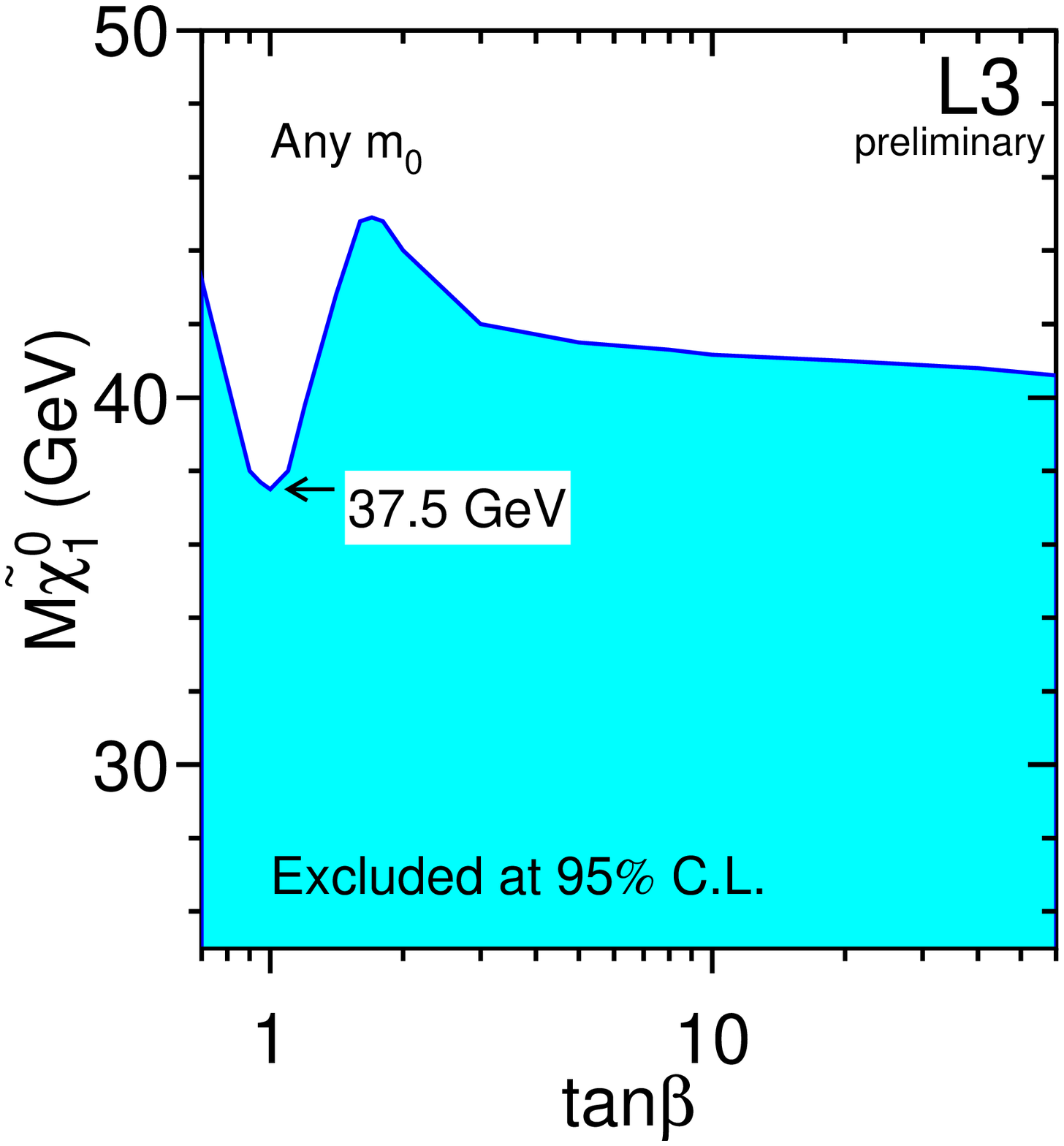} 
\end{minipage}\hspace*{10mm}
\begin{minipage}{0.45\textwidth} \includegraphics[bb=10 10 550 550,
width=7.65cm,height=6.12cm,clip=true,draft=false]{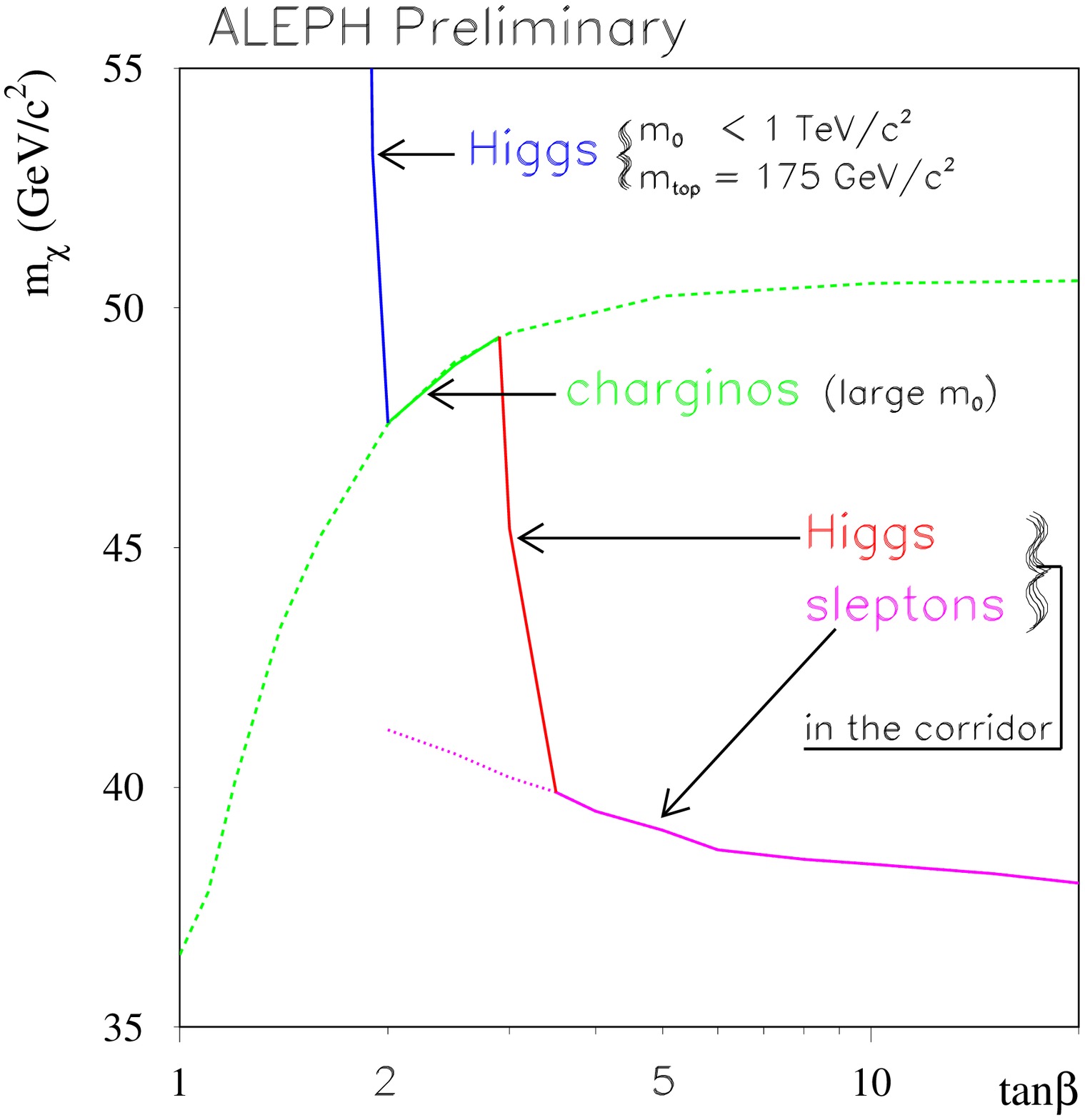} 
\end{minipage} 
\caption{95\% C.L. limits on
$\ti{\chi}_1^0$ -- LSP mass as a function of $\tan\beta$ obtained by L3 (left) 
and ALEPH (right) experiments. 
Other CMSSM parameters are scanned
in the range relevant for EW scale SUSY. ALEPH limit is obtained assuming
top quark mass of 175~GeV and $m_0<1$~TeV.}
\end{figure}

\section{Conclusions}

Productions of sleptons, squarks, charginos and
neutralinos have been searched for
at LEP2 up to $\sqrt{s}=192-202$~GeV
in various scenarios of CMSSM.  These searches cover
large variety of final states such as  leptons, lepton(s) plus
jets and jets final states all accompanied by missing energy,
ISR photon(s) accompanied by soft tracks and missing energy,
events with stable heavy charged particles or tracks with large
impact parameter.
No evidence for SUSY particle productions has been observed
and new limits have been set on their masses under specific 
assumptions on the model parameters. Model parameter independent
limits have also been derived for the light stop, lightest chargino
and lightest neutralino: $M_{\ti{t}_1}>63$~GeV, 
$M_{\ti{\chi}^{\pm}_1}>80.5$~GeV, $M_{\ti{\chi}^0_1}>37.5$~GeV.

\section*{Acknowledgements}

I am thankful  to the LEP SUSY Working Group members 
for useful discussions during the preparation of the talk. 
Special thanks 
to the organisers of the Conference for the financial support.

\end{document}